# Deep learning velocity filtering for seismic data

Xiaobin Li[1], Qiaomu Qi[1], Le Li[2], Rubing Deng[3]

1. Chengdu University of Technology, State Key Laboratory of Oil and Gas Reservoir Geology and Exploitation and Chengdu University of Technology, College of Geophysics, Chengdu 610059, China. E-mail: lixb9767@163.com; qiaomu_qi@163.com (corresponding author);

2. BGP Inc., China National Petroleum Corporation, Zhuozhou 072751 and CPNC Exploration Software Co., Ltd, Beijing 100080, China. E-mail: lile_SC@cnpc.com.cn;

3. BGP Inc., China National Petroleum Corporation, Zhuozhou 072751, China. E-mail: 1209731220@qq.com.

# ABSTRACT

Seismic velocity filtering is a critical technique in seismic exploration, designed to enhance the quality of effective signals by suppressing or eliminating interference waves. Traditional transform-domain methods, such as frequency-wavenumber (*f-k*) filtering and Radon transform filtering, often introduce aliasing and artifacts, while time-domain filtering algorithms may oversmooth effective signals. Although deep learning-based velocity filtering has yet to be directly applied in practical seismic processing, studies have demonstrated its potential in specific tasks, such as linear noise attenuation and VSP wavefield separation. However, the limited functionality of existing models poses challenges in addressing diverse velocity filtering tasks. To tackle this issue, We develop a deep learning velocity filtering algorithm based on upgoing and downgoing wave separation. By employing linear transformation, problems associated with arbitrary velocity differences are transformed in terms of upgoing and downgoing wave separation tasks. A diverse training dataset was constructed, comprising 31 theoretical velocity models, over 200 forward modeling simulations, and 46 field seismic datasets. The model trained on this dataset, combined with the proposed linear transformation strategy, achieves results comparable to traditional *f-k* and Radon filtering while also supporting tasks such as median filtering through manual wavefield picking. Applications in VSP upgoing and downgoing wavefield separation, distributed acoustic sensor (DAS) VSP P- and S-wave separation, post-stack migration arc noise mitigation, DAS common mode noise attenuation, and pre-stack CDP gather optimization demonstrate the method's versatility and practicality. This approach offers a novel and effective solution for almost all velocity filtering tasks in seismic data processing.

INTRODUCTION

Velocity filtering is a key technique in seismic exploration. This method leverages differences in wave propagation velocity to design specific filters that suppress or remove unwanted signals, thereby enhancing effective signals (Christie et al., 1983; Duncan and Beresford, 1994). In practical seismic data processing, velocity filtering is widely applied in scenarios where signals and noise exhibit velocity differences, such as suppressing linear noise, surface waves, and multiples. Additionally, this method is used for separating upgoing and downgoing waves in vertical seismic profiling (VSP) data (Li et al., 2023). Therefore, developing high-precision and high-efficiency velocity filtering algorithms is crucial for improving seismic data processing quality.

Velocity filtering algorithms are widely used in seismic exploration, and many classical methods have been developed. Common techniques include frequency-wavenumber ($f$-$k$) domain filtering (Christie et al., 1983; Kommedal and Tjøstheim, 1989; Duncan and Beresford, 1994, 1995a) and Radon transform (RT) filtering (Moon et al., 1986; Mitchell and Kelamis, 1990; Trad et al., 2002, 2003). However, these transform-domain methods often cause artifacts and spatial aliasing, especially when the data have significant amplitude variations. In addition, $f$-$k$ filtering may cause signal distortion (March and Bailey, 1983; Duncan and Beresford, 1994), while conventional Radon transform methods fail to accurately preserve amplitude (Schonewille and Zwartjes, 2002). As a result, extensive research has focused on improving these algorithms, such as the weighted least-squares Radon transform (Yi, 2018), the modified parabolic Radon transform based on singular value decomposition (SVD) (Abbad et al., 2011), and short-time Fourier transform filtering (Kazemnia Kakhki et al., 2024). Time-domain algorithms such as median filtering (Stewart, 1985; Duncan and Beresford, 1995b) and SVD filtering (Freire and Ulrych, 1988; Franco and Musacchio, 2001) can also be regarded as special types of velocity filtering methods. Since the processing is done in the time domain, problems like spatial aliasing are avoided. These algorithms primarily rely on picking seismic events to achieve effective signal separation. This reliance results in a relatively low level of automation. However, due to their reliance on manual intervention, they offer greater flexibility in wavefield separation. Another limitation of time-domain filtering methods is that they can cause signal smoothing, which may reduce accuracy. Additionally, a migration-based filtering method (Huo et al., 2019) has recently been proposed, which can achieve results similar to velocity filtering in specific applications, such as suppressing linear noise and separating VSP wavefields (Li and Zhang, 2022). However, these methods have high computational demands.

Currently, although there has been no direct research on using deep learning for velocity filtering, numerous studies have focused on specific applications, such as linear noise attenuation (Yu et al., 2019; Cheng et al., 2024), surface wave attenuation (Kaur et al., 2020; Yuan et al., 2020; Oliveira et al., 2021; Yang et al., 2023), common mode noise attenuation in distributed acoustic sensing data (DAS) (Wang et al., 2023), multiple attenuation (Breuer et al., 2020; Fernandez et al., 2024), and separating upgoing and downgoing waves in VSP (Guo et al., 2023; Li et al., 2023; Luo et al., 2023; Tao et al., 2023; Lu et al., 2024; Meng et al., 2025; Wen et al., 2025). A trained model is usually designed to separate seismic signals based on specific velocity differences. As a result, a single model may not work well for all velocity filtering scenarios. One approach is to train multiple models with different velocity differences. However, velocity variations in seismic data are complex. Therefore, relying on a limited number of models may not perform well in diverse application scenarios. In our previous research (Li et al., 2023, 2025), we focused on separating upgoing and downgoing waves, which have opposite velocity characteristics. This clear and significantly contrasting feature aids in the construction of the dataset and the training of the model. Therefore, if any velocity difference can be simplified into upgoing/downgoing wavefield separation tasks, we can focus on training a deep learning model.

We transform the wavefield separation task in any velocity range into the upgoing/downgoing wavefield separation task by applying linear moveout (LMO) transformation and data resampling. To cover diverse application scenarios, our datasets include not only VSP data but also surface seismic data, pre-stack gathers, post-stack profiles, etc. During dataset creation, we used iterative and optimization strategies to separate upgoing and downgoing waves from the raw data and generate labels. We also applied methods including selection, stitching, and LMO to create more standardized labels. The final datasets are around 2 TB, including 31 theoretical velocity models, over 200 types of simulated data from different observation systems, and 46 sets of field data. Using these diverse datasets, we trained a deep learning model of separating upgoing and downgoing waves and, with our proposed velocity filtering strategy, developed a deep learning-based velocity filtering algorithm. Based on a fixed velocity range, the proposed algorithm can automatically achieve results similar to fan-shaped $f$-$k$ filtering or rectangular $t$-$p$ filtering windows. Based on picked wave events, it also mimics median filtering. We tested the algorithm in multiple scenarios, including VSP upgoing and downgoing wavefield separation, DAS-VSP P/S wave separation, DAS common mode noise attenuation, multiple and linear noise attenuation in CDP gathers, and arc noise suppression. Comparison of results between different methods shows that the deep learning-based velocity filtering technique is more effective and versatile in all situations.

## METHOD

Velocity filtering relies on the differences in seismic wave propagation velocity to separate effective signals from noise. We have developed a wavefield separation process that simplifies all velocity variations into the specific case of upgoing and downgoing wavefields. Figure 1 shows how two waves with different velocities are converted into upgoing (negative velocity) and downgoing (positive velocity) waves. Based on the velocity difference between the red and blue waves in Figure 1, we choose a boundary velocity that falls between the two wave velocities, represented by the black dashed line. We then apply LMO to adjust the boundary line to a horizontal position, converting the two waves into upgoing and downgoing waves. The trained upgoing/downgoing wavefield separation model is used to separate the waves. Finally, inverse LMO (InvLMO) is used to restore the signals to their original state.

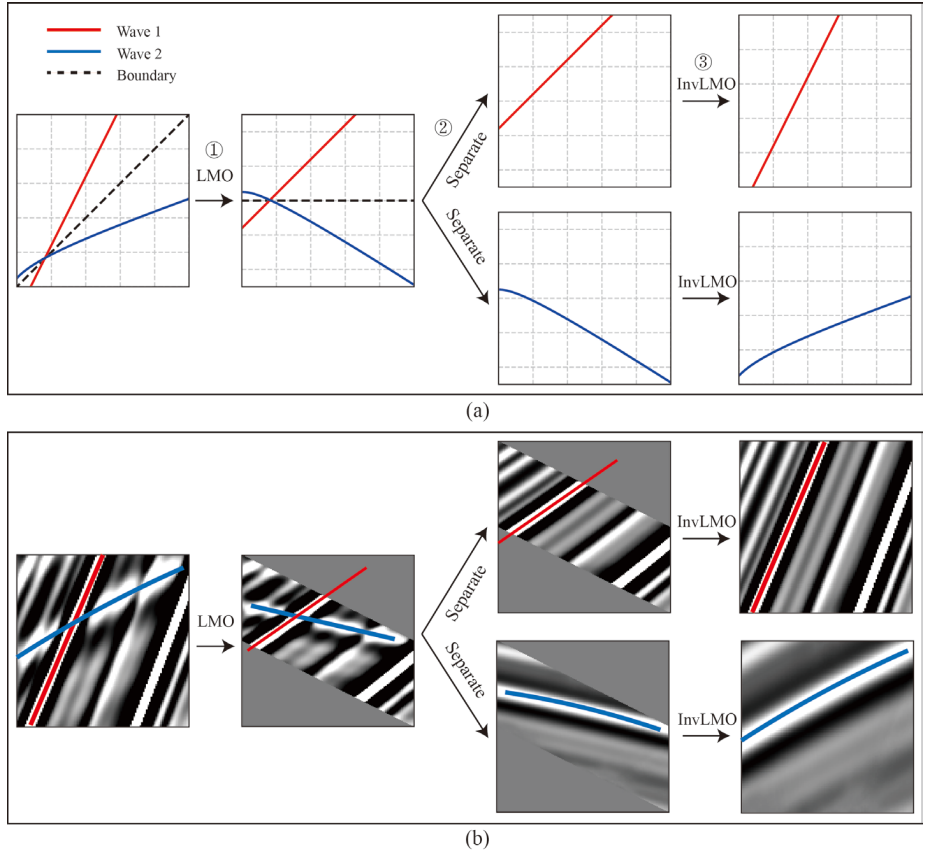

Figure 1: Flowchart of wavefield separation operation: (a) Simplified schematic, (b) Schematic of seismic data

With LMO and InvLMO, we can transform the data without loss. However, this requires that each sample point be shifted by an amount equal to a multiple of the sampling rate, which limits the ability to distinguish finer velocity differences. To simplify the process, we treat the data as having the same scale in both time and spatial directions and use slopes to represent the apparent velocities. As shown in Figure 2(a), the longitudinal direction represents time samples, and the transverse direction represents spatial samples. To achieve a lossless forward and inverse LMO, each sample point can only move in the time direction by the full grid point spacing. Therefore, the boundary slopes can only be $\pm n$, where $n$ is an integer, which limits the selection of boundary lines.

To achieve a broader range of boundary slopes, we incorporated a data resampling strategy. In the time direction, interpolation techniques are employed; in the spatial direction, equal interval extraction is used. After resampling, the calculation of the slope is as follows:

$$k = \frac{B}{IX}, I = 1,2,3...; X = 1,2,3..., \tag{1}$$

where $B$ represents the raw boundary slope, $I$ is the number of interpolations in the time direction, $X$ is the downsampling rate in the spatial direction, and $k$ represents the resampled boundary slope. The LMO and InvLMO need a time-distance (T-D) curve. Therefore the single slope value will also be transformed to the corresponding T-D curve.

Figure 2 provides examples of several slopes, with the red sample points representing the boundary T-D curve. Figure 2a shows the case with a slope of 0, used to separate positive and negative velocities. In Figure 2b, equal interval extraction is applied in the spatial direction, where $B = 1$, $I = 1$, and $X = 2$, so the resampled boundary slope is 1/2. Figure 2c demonstrates the case where interpolation is applied in the time direction, where $B = 2$, $I = 3$, and $X = 1$, so the resampled slope is 2/3. Since the calculated boundary slope needs to be converted into a T-D curve, and the T-D curve can also be manually picked, it can take any form, as shown in Figure 2d. This flexibility allows the T-D curve to be adapted to different velocity contrasts.

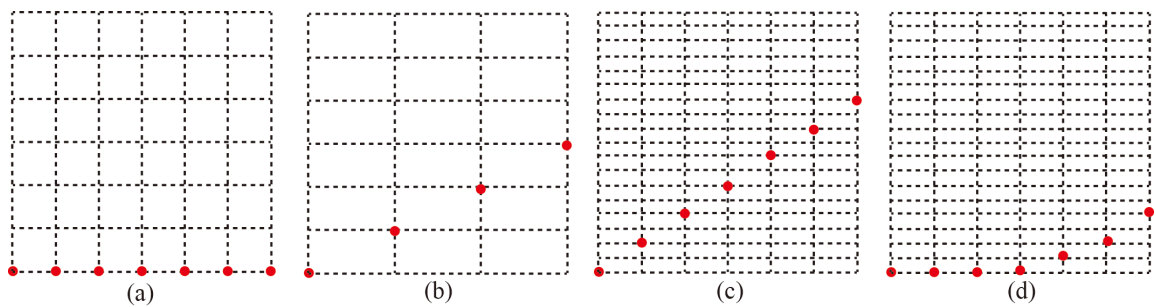

(a)  (b)  (c)  (d)

Figure 2: Schematic of data resampling and boundary T-D curve: (a) $B = 0$, $I = 1$, and $X = 1$; (b) $B = 1$, $I = 1$, and $X = 2$; (c) $B = 2$, $I = 3$, and $X = 1$; (d) Manual picking

Table 1 lists the core functions and variables used in this study. Based on application requirements, we encapsulate the core functions and provide various options, controlled by the variable "method. " Figure 3a shows the range of returned data when the "method" is set to "up" or "down." The black line represents the boundary T-D curve. When the "method" is set to "up," the wavefield with slopes less than the black solid line is returned; when the "method" is set to "down," the wavefield with slopes greater than the black solid line is returned. Figure 3b shows the data range when the "method" is set to "middle" or "outside." When the "method" is set to "middle," the wavefield between the two boundary slopes is returned; when the "method" is "outside," the wavefield outside the two boundary slopes is returned. Figures 3c and 3d illustrate the process when the "method" is "median." The "method" is named "median" because its function is similar to median filtering. First, the waveform is flattened based on the T-D curves of the picked wave events (shown in the red line in Figure 3c). Then, the flattened waveform is extracted within a limited range of positive and negative slopes. Different values for the "method" require different "boundary" inputs, as shown in Table 2.

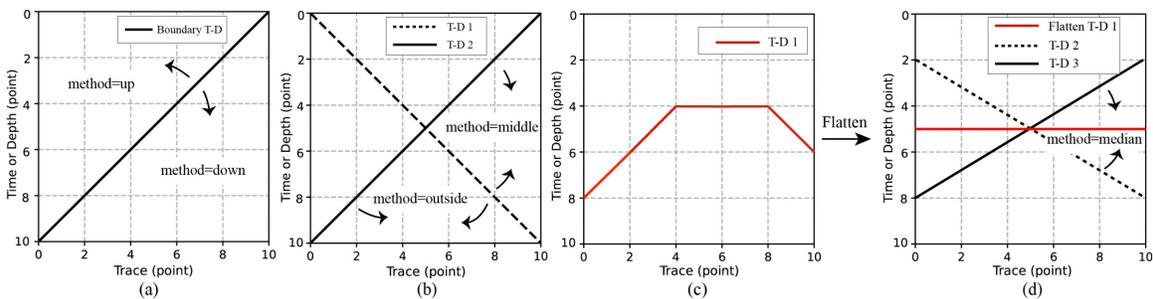

Figure 3: Schematic of the wavefield separation region: (a) "method" is set to "up" or "down"; (b) "method" is set to " middle" or " outside"; (c) "method" is set to "median",

with the red line representing the flattened T-D1 curve; (d) "method" is set to "median" and the waveform is extracted within T-D2 and T-D3

Table 1: Introduction to core function parameters

| **Function** separate(data, model, boundary=0, time_interp:int=1, sampling_x:int=1, segment_x:int =1024, segment_overlap:int =50, iter:int=1, method="up") | |
|---|---|
| data | Input seismic data |
| model | The trained deep learning model. The model output is upgoing waves |
| boundary | Boundary slope or boundary T-D curves. It is equivalent to $B$ in Equation (1) |
| time_interp | Time interpolation. It is equivalent to $I$ in Equation (1) |
| sampling_x | Spatial downsampling rate. It is equivalent to $X$ in Equation (1) |
| segment_x | Segment length in the spatial direction. The data is processed in groups of a fixed length, with each group containing "segment_x" traces. This value can be adjusted to control the GPU memory usage during the data processing, preventing memory overflow |
| segment_overlap | An overlap is set between the groups to avoid abrupt changes at the boundaries of groups. The overlapping section is replaced with the average value to ensure smooth transitions between the groups |
| iter | The number of repeated predictions |
| method | Filtering methods |

Table 2: The relationship between variables "method" and "boundary"

| method | boundary |
|---|---|
| up/down | boundary=T-D1. T-D1 can either be a single value or a one-dimensional array that matches the number of traces in the seismic data. When T-D1 is an array, it represents the input boundary T-D curve. When T-D1 is a single value, a linear T-D curve will be calculated based on the number of traces. If T-D1 is an array, the "time_interp" must be set to 1 |
| middle/outside | boundary=(T-D1, T-D2). T-D1 and T-D2 can either be single values or one-dimensional arrays. When either T-D1 or T-D2 is an array, the "time_interp" must be set to 1 |
| median | boundary=(T-D1, T-D2, T-D3). T-D1 is used to flatten the data. When either T-D2 or T-D3 is an array, the "time_interp" must be set to 1 |

DATASET

**Label optimization**

The effectiveness of the network model largely depends on the quality of the dataset. In our previous research (Li et al. 2025), we proposed an iterative strategy to gradually expand the dataset. Although the iterative strategy is generally effective, it may not always separate the data well in certain cases. Therefore, in addition to the iterative strategy, we also introduced an optimization strategy. Li et al. (2023) showed that deep learning training could overcome the limitations of the original labels, improving separation performance. This is because, during training, the network first fits the primary signals and then fits the secondary signals. This characteristic has also been utilized in some unsupervised training denoising. Therefore, we directly applied unsupervised denoising to further optimize the upgoing and downgoing wave labels. Taking the upgoing labels as an example, during the network's learning to restore the upgoing labels, the network first learns the primary feature—the upgoing waves—followed by the secondary features, such as the residual downgoing waves and some noise. Therefore, by interrupting the training at an appropriate epoch, the trained model can further refine the upgoing wave labels. We refer to this process as optimization. The network used for label optimization is the same as that used for separation. The specific network architecture is detailed in Li et al. (2023, 2025).

Figure 4 illustrates the evolution of the downgoing wave labels. After the initial separation (Figure 4b), most of the upgoing waves have been removed, but some residuals remain in certain strong events. After the iterative strategy, these strong residuals are significantly reduced, improving the quality of the downgoing wave labels. However, a small amount of upgoing wave remains. The optimization strategy further eliminates these residuals, resulting in more accurate downgoing wave labels, as shown in Figure 4d. The optimization process is more prone to damaging the effective signal, so it should be avoided when there is no obvious residual in the data.

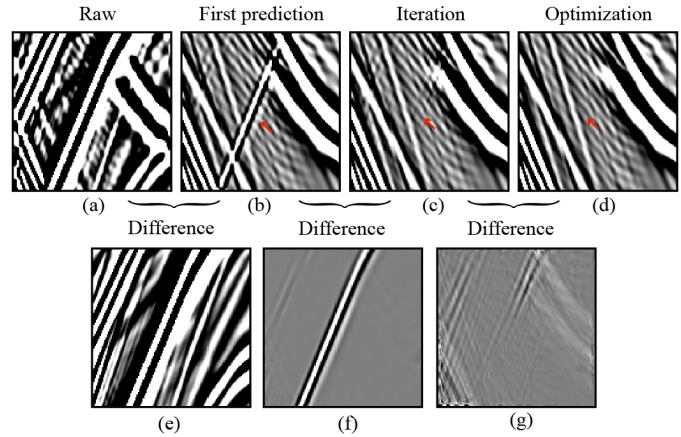

Figure 4: The evolution of the downgoing wave label during the iterative and optimization process

**Label production flow**

Our previous studies (Li et al., 2023, 2025) have primarily focused on the separation of upgoing and downgoing waves in VSP data. Here, we greatly broaden the application scenarios, which in turn requires a larger dataset. There are two main approaches for dataset creation. The first approach involves generating labels for specific tasks, such as upgoing/downgoing wavefield separation. This requires separating upgoing and downgoing waves from the raw data and improving label quality through iterative and optimization strategies. The second approach involves selecting appropriate data and using LMO to construct upgoing and downgoing wave labels, without being tied to a specific task. This approach makes better use of the collected data and expands the diversity of the dataset.

The dataset creation process for specific tasks is shown in Figure 5. First, an LMO operation is applied to the data to transform two types of waves with different apparent velocities into upgoing and downgoing waves. Then, iterative prediction is performed to separate the upgoing and downgoing waves. If the separation results are satisfactory at this stage, the separated upgoing and downgoing waves can be used as labels. Otherwise, the label optimization can be applied to remove the residuals. After optimization, if the labels remain unsatisfactory, iterative training can be used to further improve the model's adaptability to the data. In general, there are three steps for improving label quality: iterative prediction, iterative training, and label optimization. Iterative training can be further divided into two scenarios: one where training continues from the previously trained model, and another where the model is reinitialized and training begins from scratch. Table 3 ranks the effectiveness of these label refinement strategies based on trial and error results, with smaller numbers indicating better performance. A good label has fewer unwanted wave residues and retains more of the effective signal. Depending on the data characteristics and testing, the label creation strategy may not strictly follow the sequence in Figure 5 and may skip certain steps based on the specific features of the data.

Table 3: Ranking of label refinement strategies by their effectiveness, as determined through trial and error

| Strategies | Suppressing unwanted waves | Effective signal preservation |
| --- | --- | --- |
| Iterative prediction | 4 | 2 |
| Label optimization | 1 | 4 |
| Iterative training (continue training) | 3 | 1 |
| Iterative training (retraining) | 2 | 3 |

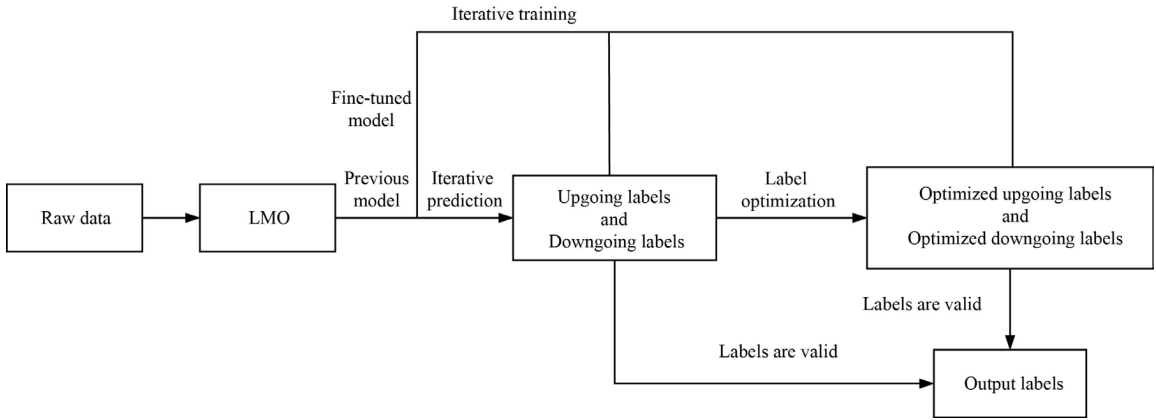

Figure 5: Dataset creation flow with specific tasks

Another type of label does not correspond to a specific task. Based on the characteristics of different data, we generate labels with upgoing and downgoing wave trends, as shown in Figure 6. First, we select a portion of the data dominated by upgoing or downgoing waves, or apply LMO to construct these waves. Figure 7 illustrates the upgoing and downgoing waves generated in these two cases. From Figure 7, we can see that the constructed wavefields usually have one dominant wave type, but some unwanted waves remain in the data. Therefore, these data need to be separated to obtain more accurate upgoing and downgoing wave labels. Next, the labels are screened. A signal-to-noise ratio (S/N) threshold is set to assess the similarity between the predicted and original labels. Typically, a higher S/N is aimed at ensuring that the generated wave is dominated by the target wave, reducing reliance on the separation accuracy. If the labels still contain noticeable interference or noise, label optimization strategies can be applied to further refine and improve the labels.

By constructing upgoing and downgoing waves, we can obtain higher-quality labels. This is because we can select and create signals that mainly contain the target wave, and in some cases, directly obtain seismic signals with only the target wave. As a result, these labels rely less on iterative and optimization strategies. Additionally, we can make full use of the collected data, avoiding problems with insufficient data for certain types. For example, with post-stack seismic data, it can be difficult to create a large number of labels for specific seismic processing tasks. However, by using LMO to construct upgoing and downgoing waves that match the post-stack waveforms, we can effectively supplement these data.

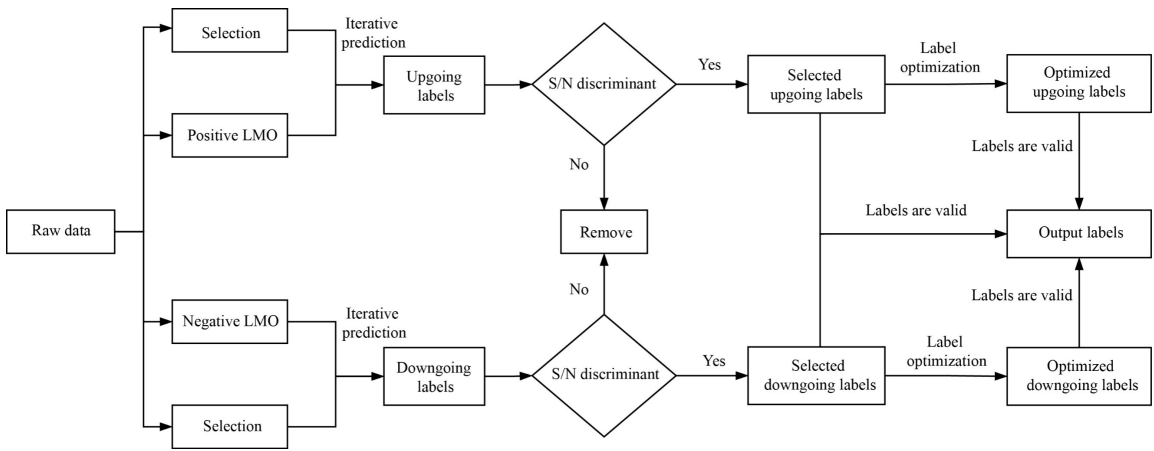

Figure 6: Dataset creation flow without specific tasks

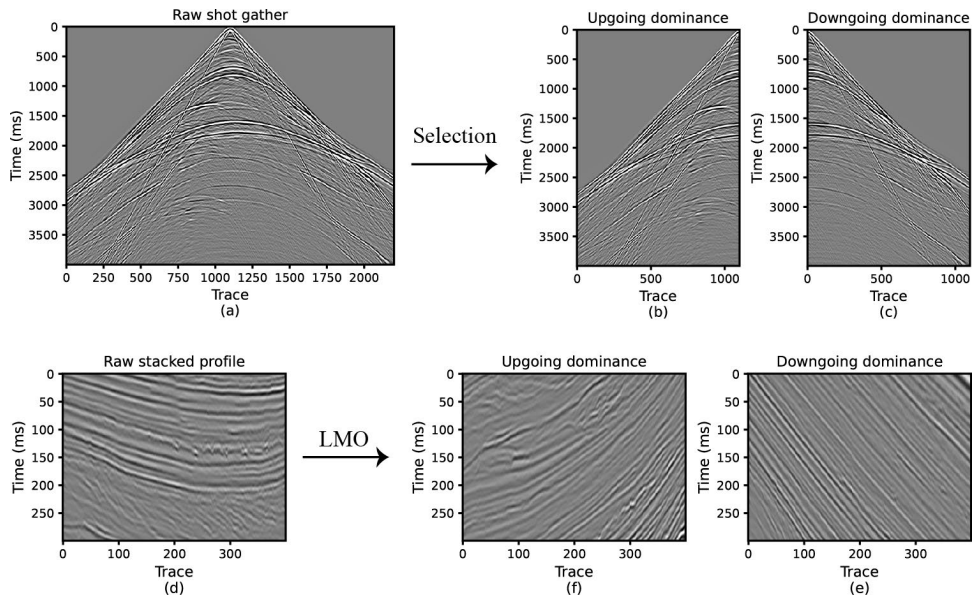

Figure 7: Schematic of generated upgoing and downgoing waves. (a) Original shot gather data; (b) Upgoing wave-dominated shot gather; (c) Downgoing wave-dominated shot gather; (d) Post-stack profile; (e) Upgoing wave signal from LMO operation; (f) Downgoing wave signal from LMO operation

**Training and validation datasets**

We conducted forward modeling for various data types, including VSP, surface observation systems, acoustic waves, and elastic waves. Additionally, we collected several types of field data, such as VSP, surface seismic, pre-stack gathers, and post-stack profiles. After processing these data, we generated numerous labels. The simulated data includes 31 velocity models, and accounting for differences in observation systems and modeling parameters, it encompasses over 200 distinct simulated datasets. For field data, we collected 46 sets. During the label creation process for each dataset, we also performed resampling to expand the dataset. Table 4 provides the details of the complete dataset. More information can be found at: https://geolxb.github.io/openWFS/datasets/. In these datasets, 5 sets were selected for the validation set, with the remaining data used for training. The dataset includes two sizes: 128×128 and 128×48. This is because, in some application scenarios (e.g., imaging pre-stack gathers), the data often contains fewer than 128 traces per common depth point (CDP), making 128×128 labels unfeasible. After testing, we found that larger labels performed better. Therefore, depending on the data feature, we used both sizes for label creation.

Table 4: Dataset Details

| Data type | | Patch size | Number |
|---|---|---|---|
| Convolutional seismic data | VSP (Li et al., 2025) | 128×128 | 822448×2 |
| Wave equation simulation | VSP | 128×128 | 4233881×2 |
| | Surface | 128×128 | 4019982×2 |
| Field data | VSP | 128×128 | 3892200×2 |
| | | 128×48 | 453561×2 |
| | Surface | 128×128 | 84567×2 |
| | | 128×48 | 2339116×2 |
| | CMP/CRP/CDP | 128×48 | 4962676×2 |
| | post-stack | 128×128 | 2215242×2 |

**Test datasets**

To more systematically evaluate the performance of the trained model, we constructed three test datasets, each including both original amplitude and amplitude-balanced cases. The original amplitude is used to assess the model's separation accuracy under varying energy conditions, particularly its performance for strong events. In practical applications, energy-compensated data is often used for wavefield separation; therefore, we also incorporated automatic gain control (AGC) in the test data to provide a more comprehensive assessment of the model's performance. We use four metrics, the S/N, peak signal-to-noise ratio (PS/N), coefficient of determination (R2), and structural similarity index measure (SSIM) for quantitative analysis. The corresponding formulas can be found in Li et al. (2025).

The first and second test datasets are the same as those used in Li et al. (2025). The first test dataset uses the simulation method proposed by Margrave and Daley (2014), which simulates separated VSP upgoing and downgoing waves. This method is only applicable to zero-offset VSP data. We applied different wavelets and used various velocity logs and attenuation parameters to simulate data for 100 shots. The test data is shown in Figure 8.

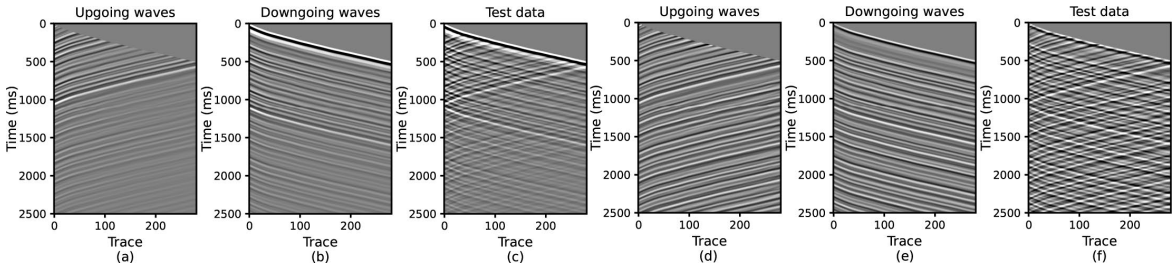

Figure 8: Test data 1. (a), (b), and (c) show the upgoing wave, downgoing wave, and unseparated wavefield with original amplitudes, respectively. (d), (e), and (f) show the upgoing wave, downgoing wave, and unseparated wavefield after AGC, respectively

The second test dataset is presented in Figure 9. It is simulated using a nearly horizontal layered medium and is created by segmenting and stitching the original data. The detailed creation process can be found in Li et al. (2025).

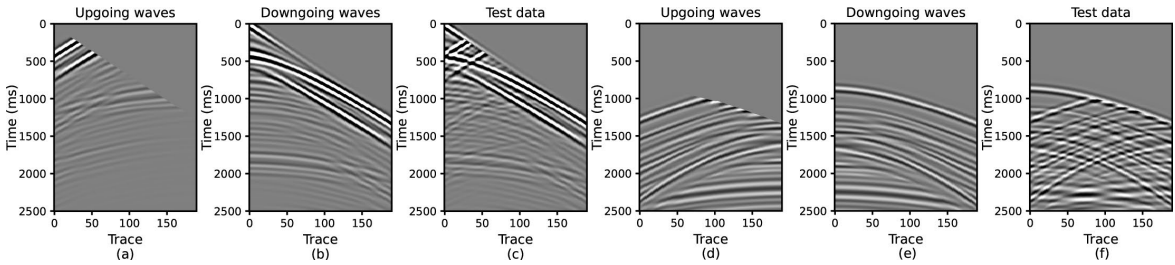

Figure 9: Test data 2. The subplots have the same meanings as those in Figure 8

The third test dataset is similar to the second one, with the main difference being that the upgoing wave has not been muted, simulating linear noise that runs through the effective signals, as shown in Figure 10. Additionally, during the simulation, the surface boundary is set as a free boundary, making the wavefield more complex compared to the second test dataset.

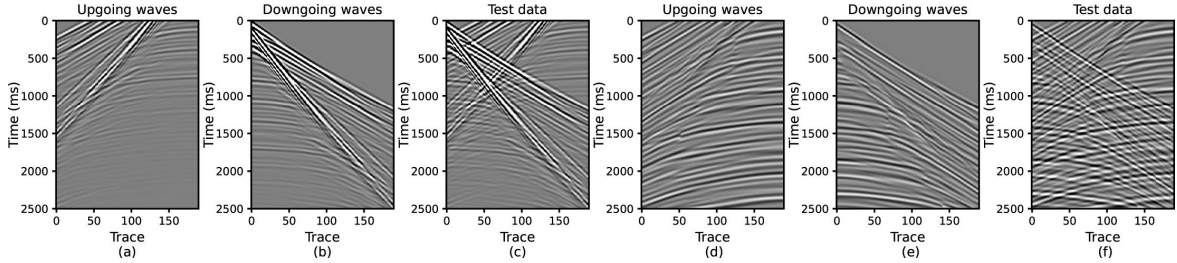

Figure 10: Test data 3. The subplots have the same meanings as those in Figure 8

## TRAINING AND TEST

**Training**

In our previous studies, we found that training with the upgoing wave as the label led to better overall separation performance. Considering the computational cost, we only trained a model with the upgoing wave as the label. The downgoing wave is obtained by subtracting the upgoing wave from the original data. To accommodate a wider range of application scenarios and expand the dataset, we performed data augmentation during the training phase. The specific data augmentation process is shown in Figure 11. First, 500,000 patch pairs (matched upgoing and downgoing wave patches) of size 128×128 and 200,000 patch pairs of size 128×48 are randomly selected from the entire dataset. The patch pairs are then divided into three categories based on a fixed probability: 40% use the original patch pairs, 40% randomly match upgoing and downgoing patches, and 20% use only upgoing or downgoing patches. Within each category, there is a probability of deciding whether to use the original patches or flipped patches. In single-patch training, if the input is an upgoing patch, the output will be the same as the input. If the input is a downgoing patch, the label will be a zero-value patch. This operation accounts for the fact that not all data require wavefield separation during processing; some data may only have local wavefield mixing. Data flipping includes three cases: time axis flipping, spatial axis flipping, and flipping both axes simultaneously, with equal probabilities for each.

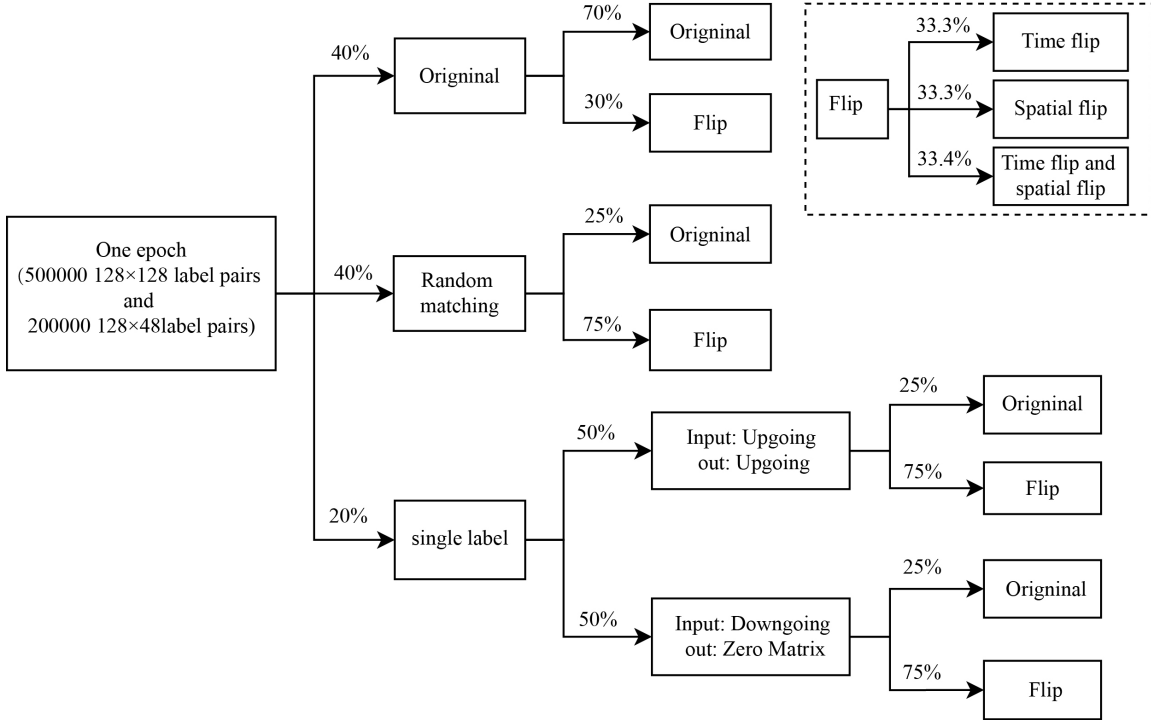

Figure 11: Data augmentation

In addition to data augmentation, we also modified the data normalization method and the loss function. In previous studies, maximum value normalization was applied to the data based on gathers. However, in this study, we switched to patch-level normalization, aiming to balance the contribution of each patch and prevent the model from favoring patches with larger amplitudes during training. As for the loss function, we replaced the L1 loss function with the square root of the L1 (sqrt L1) loss function. This adjustment is intended to balance the contributions of samples within one patch, reducing the model's over-reliance on high-amplitude events while prioritizing overall fitting accuracy.

$$sqrtL1 = \sqrt{\|\mathbf{label} - \mathbf{predict}\|_1} \qquad (2)$$

where **label** is the label of datasets and **predict** represents the network's output.

Figure 12 presents the test results for these two modifications. The S/N was calculated based on the validation set. To ensure a fair comparison, data normalization in the validation set was performed at the gather level, meaning patch normalization was applied only to the training set. As shown in Figure 12, both the use of sqrt L1 and patch normalization resulted in a significant improvement in the S/N. Among these, sqrt L1 made the greatest contribution, as it not only balances amplitude values across different patches but also amplitude values within one patch.

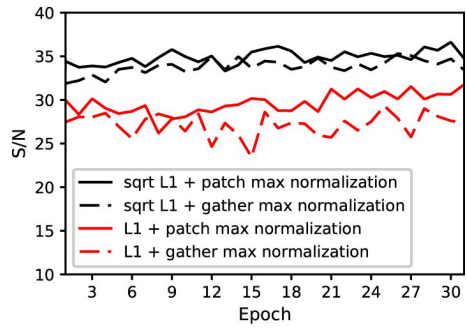

Figure 12: Comparison of S/N under different training strategies

We trained the datasets using a batch size of 64 and an initial learning rate of 0.0001, with the model's initial parameters taken from the final training results of our previous research (Li et al. 2025). The variations of loss value and S/N during training are shown in Figure 13. After 1000 epochs of training, the S/N of the validation set stabilized around 39.

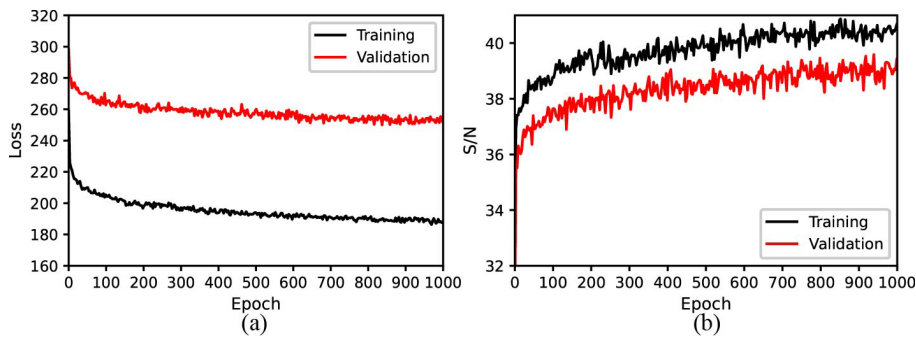

Figure 13: The variations of (a) loss and (b) S/N with epochs

**Test**

We compared traditional methods, including *f-k* filtering, frequency-domain Radon transform (FDRT), and sparse inversion Radon transform (SIRT). Table 5 summarizes all the test results and demonstrates that our latest trained model outperforms all other methods in nearly every metric. To facilitate comparison, we calculated the average of all metrics for each dataset. Since the S/N and PS/N do not lie within the 0-1 range, we normalized these values. Figure 14 presents the overall metric results for each test data. From a quantitative perspective, the model trained in this study performed well, clearly outperforming both the traditional methods and our previous research (Li et al., 2025). The improvement was particularly notable for data without AGC, indicating that the trained model has enhanced its ability to handle strong events.

Table 5: The comparison of test metrics

| Test data | Type | | *f-k* | FDRT | SIRT | DL-Previous model (Li et al., 2025) | DL-Latest model |
|---|---|---|---|---|---|---|---|
| Data 1 | S/N | Up | 15.89 | 9.92 | 16.55 | 25.22 | 27.99 |
| | | Down | 24.10 | 8.66 | 15.22 | 33.43 | 36.20 |
| | PS/N | Up | 34.76 | 28.80 | 35.42 | 44.10 | 46.86 |
| | | Down | 50.17 | 34.73 | 41.28 | 59.50 | 62.27 |
| | SSIM | Up | 0.9849 | 0.8831 | 0.9052 | 0.9958 | 0.9970 |
| | | Down | 0.9984 | 0.9568 | 0.9675 | 0.9997 | 0.9998 |
| | R2 | Up | 0.9721 | 0.8954 | 0.9748 | 0.9968 | 0.9983 |
| | | Down | 0.9960 | 0.8466 | 0.9429 | 0.9995 | 0.9997 |
| | Model epoch | | - | - | - | - | 1000 |
| Data 1-AGC | S/N | Up | 22.55 | 7.99 | 15.98 | 28.40 | 29.13 |
| | | Down | 22.77 | 11.90 | 21.13 | 28.62 | 29.36 |
| | PS/N | Up | 35.58 | 21.02 | 29.02 | 41.44 | 42.17 |
| | | Down | 35.60 | 24.73 | 33.96 | 41.46 | 42.19 |
| | SSIM | Up | 0.9837 | 0.8003 | 0.9292 | 0.9943 | 0.9956 |
| | | Down | 0.9840 | 0.8816 | 0.9564 | 0.9942 | 0.9953 |
| | R2 | Up | 0.9943 | 0.8379 | 0.9741 | 0.9985 | 0.9987 |
| | | Down | 0.9946 | 0.9287 | 0.9916 | 0.9986 | 0.9988 |
|  | Model epoch | | - | - | - | - | 1000 |
| Data 2 | S/N | Up | 4.69 | 4.28 | 6.70 | 15.00 | 20.72 |
|  |  | Down | 15.32 | 10.00 | 16.53 | 25.62 | 31.34 |
|  | PS/N | Up | 27.22 | 26.81 | 29.23 | 37.53 | 43.25 |
|  |  | Down | 36.88 | 31.56 | 38.10 | 47.19 | 52.91 |
|  | SSIM | Up | 0.9133 | 0.8872 | 0.8732 | 0.9604 | 0.9853 |
|  |  | Down | 0.9892 | 0.9670 | 0.9758 | 0.9965 | 0.9989 |
|  | R2 | Up | 0.5551 | 0.5461 | 0.7134 | 0.9480 | 0.9837 |
|  |  | Down | 0.9703 | 0.8990 | 0.9759 | 0.9971 | 0.9992 |
|  | Model epoch | | - | - | - | - | 1000 |
| Data 2-AGC | S/N | Up | 10.75 | 6.99 | 10.56 | 15.45 | 17.56 |
|  |  | Down | 10.85 | 8.63 | 11.92 | 15.55 | 17.65 |
|  | PS/N | Up | 26.82 | 23.05 | 26.62 | 31.51 | 33.62 |
|  |  | Down | 21.48 | 19.27 | 22.55 | 26.18 | 28.29 |
|  | SSIM | Up | 0.8604 | 0.7572 | 0.8142 | 0.9408 | 0.9568 |
|  |  | Down | 0.8477 | 0.7610 | 0.7814 | 0.9174 | 0.9402 |
|  | R2 | Up | 0.9156 | 0.7989 | 0.9115 | 0.9711 | 0.9821 |
|  |  | Down | 0.9175 | 0.8626 | 0.9353 | 0.9717 | 0.9824 |
|  | Model epoch | | - | - | - | - | 1000 |
| Data 3 | S/N | Up | 14.43 | 8.18 | 15.59 | 20.09 | 30.15 |
|  |  | Down | 17.24 | 8.72 | 17.98 | 22.90 | 32.95 |
|  | PS/N | Up | 34.65 | 28.39 | 35.80 | 40.31 | 50.36 |
|  |  | Down | 38.31 | 29.80 | 39.06 | 43.98 | 54.03 |
|  | SSIM | Up | 0.9623 | 0.9265 | 0.9561 | 0.9865 | 0.9967 |
|  |  | Down | 0.9780 | 0.9446 | 0.9716 | 0.9912 | 0.9990 |
|  | R2 | Up | 0.9393 | 0.8432 | 0.9559 | 0.9834 | 0.9975 |
|  |  | Down | 0.9780 | 0.8655 | 0.9818 | 0.9946 | 0.9994 |
|  | Model epoch | | - | - | - | - | 1000 |
| Data 3-AGC | S/N | Up | 12.59 | 7.50 | 12.39 | 18.27 | 21.25 |
|  |  | Down | 11.78 | 8.12 | 13.16 | 17.45 | 20.43 |
|  | PS/N | Up | 23.56 | 18.47 | 23.35 | 29.23 | 32.21 |
|  |  | Down | 22.77 | 19.10 | 24.15 | 28.44 | 31.42 |
|  | SSIM | Up | 0.8650 | 0.7324 | 0.8314 | 0.9435 | 0.9637 |
|  |  | Down | 0.8104 | 0.7116 | 0.7792 | 0.8981 | 0.9528 |
|  | R2 | Up | 0.9439 | 0.8220 | 0.9416 | 0.9849 | 0.9923 |
|  |  | Down | 0.9323 | 0.8450 | 0.9508 | 0.9817 | 0.9907 |
|  | Model epoch | | - | - | - | - | 1000 |

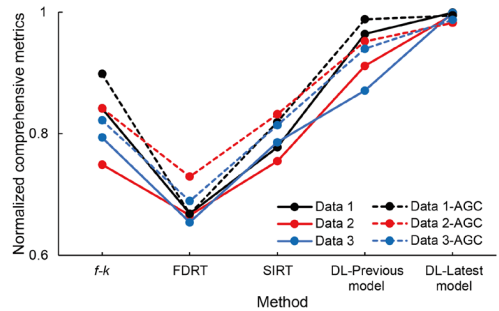

Figure 14: Normalized comprehensive metrics

EXAMPLES

**Case 1: DAS-VSP wavefield separation**

Currently, most DAS-VSP data is still single-component. In this case, conventional multi-component techniques cannot be directly applied. Therefore, in practical applications, we can separate P- and S-waves based on their slope differences. In near-offset data, the slopes of P-waves and S-waves are relatively fixed, while in far-offset data, these slopes differ significantly from those in near-offset data. In such cases, we rely on manually calibrated T-D curves for P- and S-waves and then perform wavefield separation. The raw data is shown in Figure 15. We extracted the T-D curves for four types of waves and sequentially separated them in the following order: downgoing-P, downgoing-S, upgoing-P, and upgoing-S. The separation parameters used are as follows:

Table 6: Application parameters for case 1

| parameters | boundary | sampling_x | time_interp | method |
|---|---|---|---|---|
| value | (T-D,1,-1) | 1 | 1 | median |

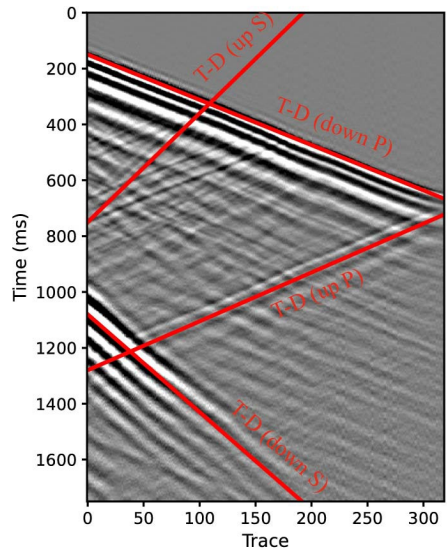

Figure 15: Raw DAS-VSP data

The process is similar to median filtering, so we also applied median filtering to sequentially extract the four types of waves. The T-D curves used in both methods are the same. The separation results are shown in Figure 16. In median filtering, the size of the spatial window is a key parameter. We tested two different window sizes: a smaller window (30 traces) and a larger window (60 traces). When the spatial window is small, the wavefield cannot be effectively separated, as shown in Figure 16a (indicated by the red circle), where strong S-waves are residual in the separated P-waves. When the spatial window is larger, the separated wavefield is relatively cleaner, but the larger window causes data smoothing, resulting in more residual signals, as shown in Figure 16j, where a significant amount of effective signal remains in the residual wavefield. In contrast, the deep learning separation method we proposed achieves better wavefield separation. The residual wavefield primarily exhibits horizontal common mode noise and some signals with larger slopes, which are distinct from the slopes of the four types of waves.

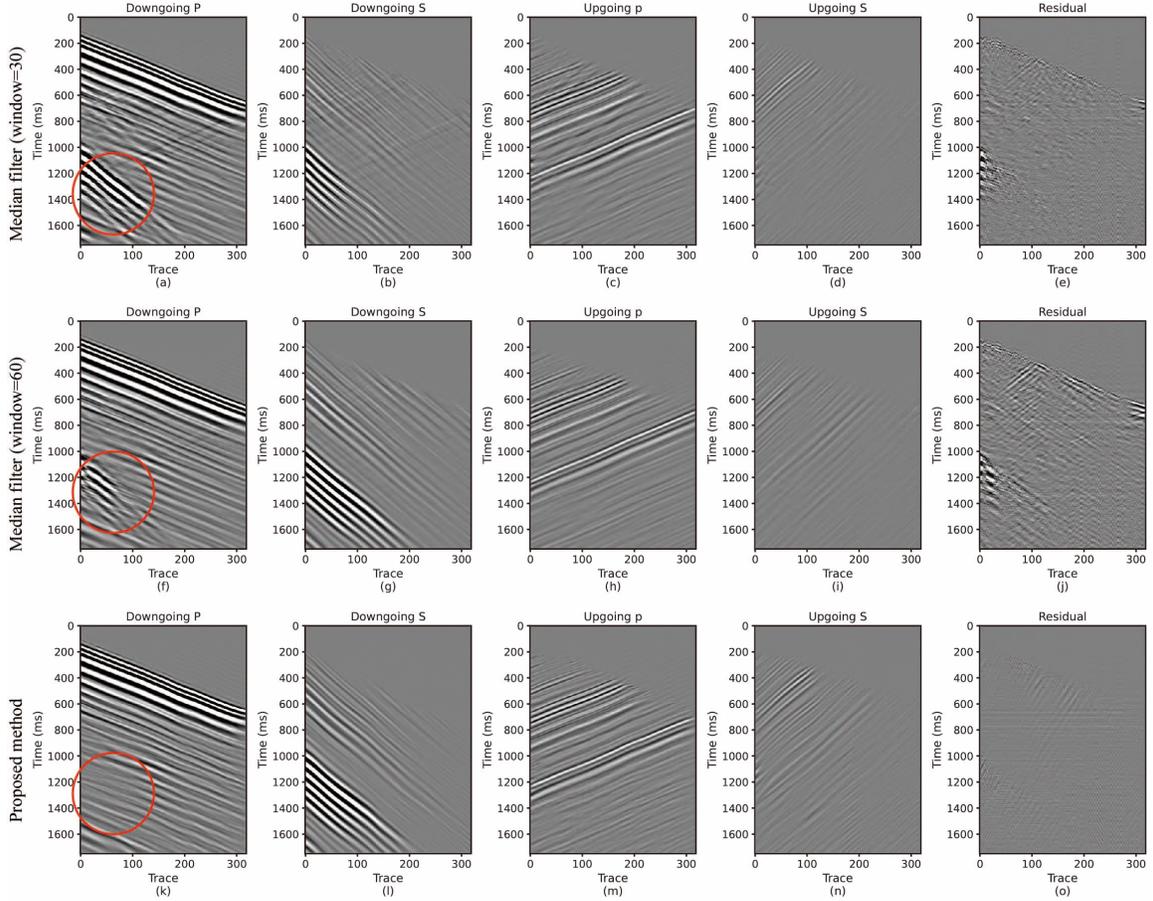

Figure 16: Comparison of wavefield separation methods for DAS-VSP data. (a) to (e) show the small window median filtering results, including downgoing P-waves (a), downgoing S-waves (b), upgoing P-waves (c), upgoing S-waves (d), and the residual wavefield (e). (f) to (j) show the large window median filtering results, including downgoing P-waves (f), downgoing S-waves (g), upgoing P-waves (h), upgoing S-waves (i), and the residual wavefield (j). (k) to (o) show the results from the proposed deep learning method, including downgoing P-waves (k), downgoing S-waves (l), upgoing P-waves (m), upgoing S-waves (n), and the residual wavefield (o)

**Case 2: Migration arc noise**

In seismic data imaging, complex structures, uneven coverage, and imperfections in imaging methods can lead to large-angle arc-shaped noise in the post-imaging data. Figure 17 illustrates an example, with the red line highlighting these noisy components. If seismic horizon data is available, the main signals can be extracted along the trend of seismic horizons. However, even without seismic horizon data, the proposed method remains effective in removing large-angle noise when the subsurface structure is relatively simple. In this instance, where the seismic horizons are nearly horizontal, we directly suppress the large-angle noise. The processing parameters are as follows:

Table 7: Application parameters for case 2

| parameters | boundary | sampling_x | time_interp | method |
|---|---|---|---|---|
| value | (4,-4) | 1 | 7 | middle |

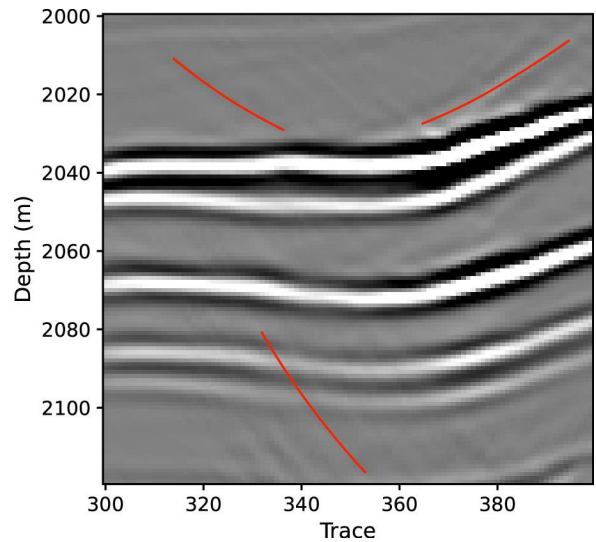

Figure 17: Arc noise in imaging

We used *f-k* filtering as a comparison method to evaluate the filtering effect. Two fan-shaped filter windows were selected: a smaller window to suppress noise more effectively, and a larger window to preserve as much effective signal as possible. The separation results are shown in Figure 18. From these results, both the smaller and larger *f-k* filter windows led to a loss of effective signals, as highlighted by the red lines in Figure 18. The red arrows point to the arc noise that *f-k* filtering failed to remove. In contrast, the deep learning filtering method preserved more effective signals and provided better noise suppression.

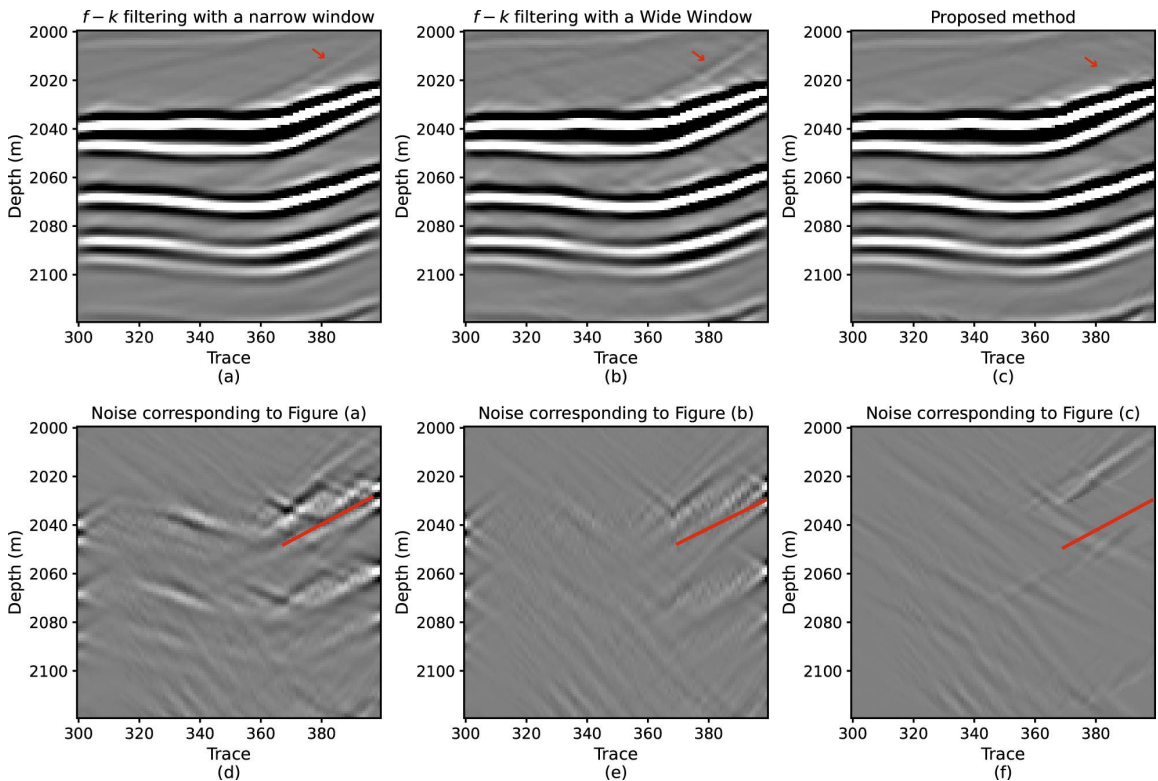

Figure 18: Arc noise suppression results. (a) and (d) show the effective signal and removed noise using the *f-k* filtering with a narrow filter window; (b) and (e) show the effective signal and removed noise using the *f-k* filtering with a wide filter window; (c) and (f) show the effective signal and removed noise using the proposed deep learning method

**Case 3: DAS common mode noise**

In DAS data, common mode noise often appears as horizontal events, and each trace of common mode noise remains consistent. This type of noise can be suppressed using a simple stacking method (Willis, 2022). This method statistically estimates the noise distribution and is usually effective in simpler cases. However, in more complex field environments, the data may be affected by various interferences, or the pattern of common mode noise may be disrupted during data processing, leading to the failure of the stacking method to fully remove such noise. In such cases, velocity filtering can be applied to suppress the noise. Figure 19 presents an example, with the processing parameters for this data as follows:

Table 8: Application parameters for case 3

| parameters | boundary | sampling_x | time_interp | method |
|---|---|---|---|---|
| value | (1,-1) | 2 | 1 | outside |

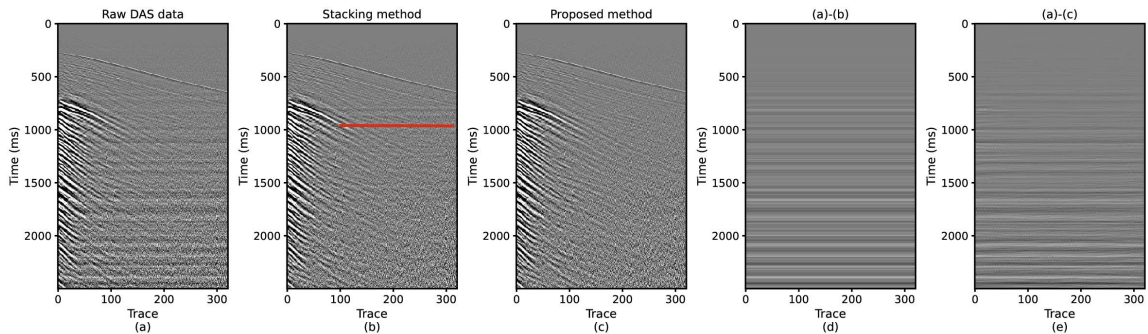

Figure 19: Common mode noise suppression results: (a) the raw data; (b) the result of the stacking method; (c) the result of the proposed method; (d) the noise suppressed by the stacking method; and (e) the noise suppressed by the proposed method

Using the stacking method, most of the noise is suppressed, as shown in Figure 19b. However, due to significant environmental noise and strong S-wave energy, the statistical noise characteristics are not accurate enough, resulting in incomplete suppression of the common mode noise. The red line in Figure 19 highlights the residual noise. In this case, the deep learning filtering method is more effective in suppressing common mode noise, with almost no noticeable remaining noise, as shown in Figure 19c.

**Case 4: Common depth point (CDP) gather optimization**

In the CDP gather, there is typically a large amount of noise, such as linear noise and multiple reflections. These types of noise differ significantly from the effective signal in terms of their apparent velocities. When the imaging velocity model is accurate, the effective signal is often flattened, while noises such as multiples appear in a parabolic shape. With sufficient coverage, these noises can be well suppressed through stacking. However, they can significantly impact tasks like pre-stack inversion and AVO (amplitude versus offset) feature extraction. Since multiple reflections often exhibit a parabolic event in the CDP gather, the parabolic Radon transform method is widely used. In this case, we used the weighted least squares parabolic Radon transform (WLSRT) as a comparison method. Figure 20 shows the denoising results. For the parabolic Radon transform, one important parameter is the maximum time shift at the far offset. Based on data sampling and offset range, the maximum time shift was set to 71ms. The deep learning parameter settings are as follows:

Table 9: Application parameters for case 4

| parameters | boundary | sampling_x | time_interp | method |
|---|---|---|---|---|
| value | (1,-1) | 1 | 2 | middle |

The data consists of 72 traces with a time sampling rate of 2 ms. Based on the parameters mentioned above in Table 9, the equivalent maximum time shift is also 71 ms.

Therefore, the maximum time shift for both the deep learning filtering and the WLSRT was set to 71ms. In Figure 20, the WLSRT method shows significant energy enhancement at the data boundaries (as shown in Figure 20b). This is a common drawback of transform domain methods, where the performance is often poorer at the data boundaries.

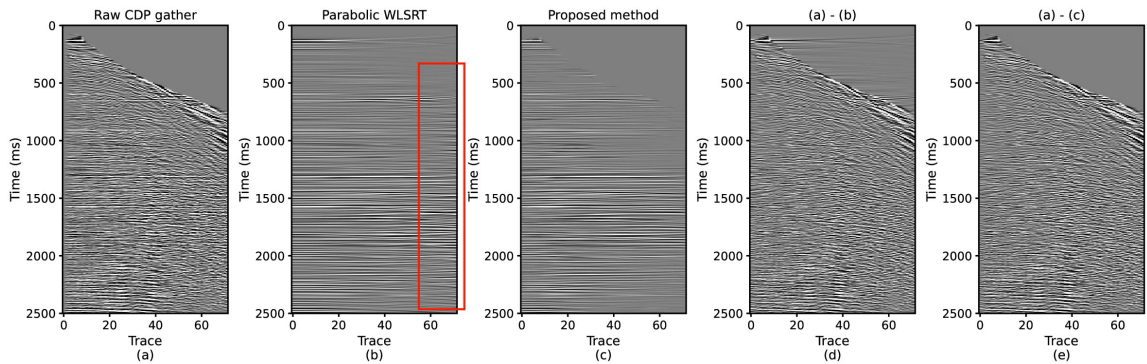

Figure 20: CDP gather optimization. (a) raw data; (b) result of WLSRT; (c) result of the proposed method; (d) noise suppressed by WLSRT; (e) noise suppressed by the proposed method

As mentioned above, through seismic data with multiple coverage, stacked profiles can also effectively suppress noise. In this case, the coverage is sufficient, and if the goal is to obtain the final stacked profile, there is no need to heavily optimize the CDP gather data. To further compare the effectiveness of methods, we performed stacking on the denoised data. Figures 21a to 21c show the stacking results of the raw data, the WLSRT, and the proposed method, respectively. From these three stacking results, there is no obvious difference. Although traditional methods may produce more aliasing and artifacts, these artifacts and aliasing typically lack coherence, and after stacking, they are effectively suppressed. However, the stacking process may also amplify the loss of effective signals. Although there is no obvious loss of effective signals in Figures 20d and 20e, after stacking, as shown in Figures 21d and 21e, some loss of effective signals can be observed. The loss of wave events in Figure 21e is less compared to Figure 21d. Figure 22 shows the frequency spectrum of the data in Figure 21. Overall, the spectral difference between the two methods in the stacked profile shown in Figure 22a is minimal. In Figure 22b, it can be observed that the deep learning method mainly loses low-frequency signals, while the WLSRT method mainly loses mid-to-high-frequency signals. Another significant difference between the two methods is their computational efficiency. The inversion-based Radon transform method typically requires iterative computations, which results in lower efficiency. In contrast, our proposed method, once adequately trained, can be directly applied, offering higher efficiency.

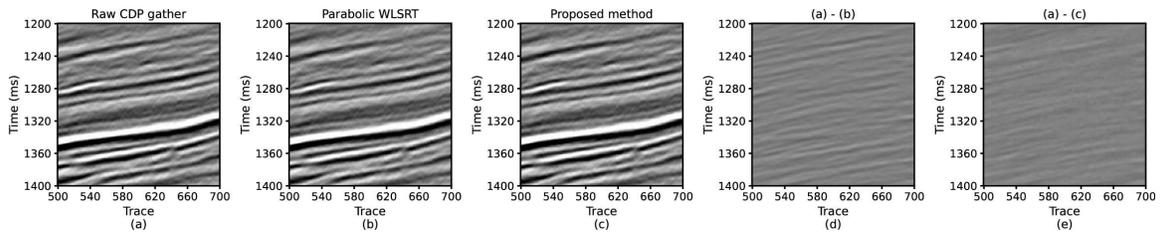

Figure 21: Stacked profiles. (a) raw data; (b) WLSRT; (c) proposed method; (d) residual signal by WLSRT; (e) residual signal by the proposed method

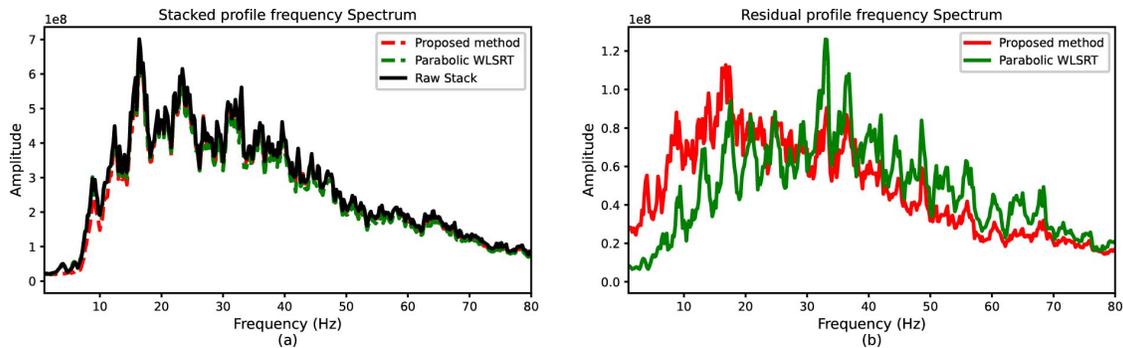

Figure 22: Frequency spectrum of stacked profiles. (a) effective signal stacked profiles; (b) residual signal stacked profiles

DISCUSSION

**Parameter suggestions**

We provide a Python function for deep learning velocity filtering, which includes multiple adjustable parameters. Some recommendations for parameter adjustments are provided in Table 10. The parameter "boundary" directly controls the slope (apparent velocity) and should be adjusted first. If the "boundary" is based on the picking T-D curve, there is typically no need to focus much on auxiliary parameters like "time_interp" and "sampling_x." However, when using a fixed slope value, setting the "boundary" to an integer usually yields better results. When the "boundary" be set to a non-integer value. In this case, the shift value in the T-D curve may become a decimal, but only integer shifts are allowed. As a result, these decimals will be forced to convert to integers, which may create a step structure in the T-D curve. Therefore, slopes with an absolute value smaller than 1, except for zero, should be avoided whenever possible. The slope can also be controlled using the parameter "time_interp." For depth-domain data with large sampling rates, fine control over the slope is difficult, so adjustments to the "time_interp" are necessary. The parameter "sampling_x," with a default value set to one, usually does not require adjustment. However, for DAS-VSP data with high spatial density (e.g., a spatial sampling rate of 0.5 m), adjusting "sampling_x" can help control the slope.

Table 10: Parameter recommendations for Python functions

| Parameters | Adjust priority | Default Value | Recommendation |
|---|---|---|---|
| boundary | 1 | 0 | If a manually picked T-D curve is available, it should be used first; otherwise, integer values are preferred. Absolute values less than 1 may introduce a horizontal step structure in the T-D curve |
| time_interp | 2 | 1 | It must be a positive integer and should be used when the time sampling rate is insufficient |
| sampling_x | 3 | 1 | It must be a positive integer and should be used when the spatial sampling rate is sufficiently high |

**Anomalous amplitude influence**

In practical applications, separation failures are common. For simulated data, most failures are caused by excessively strong energy and prominent wave events. These can be addressed using the LMO and stitching methods to create labels with strong energy events. However, in field data, in addition to strong energy events, noise—particularly anomalous amplitude noise—can also be present. In simple terms, strong energy, whether from noise or seismic signals, may cause issues in deep learning filtering. Therefore, it is recommended to first suppress anomalous amplitudes and strong energy before applying deep learning filtering.

**Considerations for label refinement**

In addition to removing unwanted waves, label optimization can also reduce noise in the labels. Early stopping during iterative training can also help with noise reduction. However, training with completely noise-free labels may limit the model's ability to adapt to field data, as field data often contains various types of noise. Moreover, a certain level of noise in the labels can improve the model's generalization.

We created two training sets. One set had labels that were heavily optimized to remove most of the noise, making the waveforms cleaner. The other set had labels with only moderate optimization, so it included more random noise. Both training sets shared the same validation set, which was optimized and did not contain significant noise. As shown in Figure 23, the model trained on a dataset with some noise achieved a higher S/N on the validation set, indicating better generalization performance. Therefore, For labels without clear upgoing and downgoing wave mixing, the label optimization strategy should be avoided. Allowing some random noise to remain can improve the model's ability to generalize.

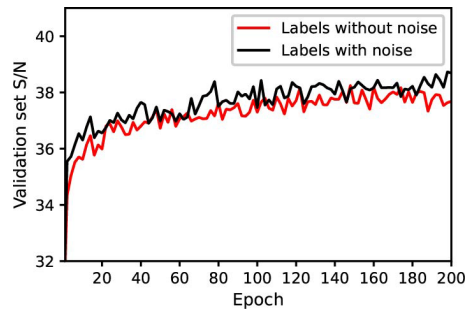

Figure 23: S/N Comparison of the validation set. The black line represents validation results from the model trained on a dataset that retains noise from field data, while the red line shows the validation results from a dataset that has been over-optimized to remove noise

CONCLUSIONS

Based on the features of transfer learning and unsupervised training, we developed strategies such as iteration and optimization to refine labels. For task-specific data, we extracted labels from the original dataset. Moreover, we proactively created labels with similar upgoing and downgoing wave features, which were not originally present, thereby generating a more diverse and high-quality dataset. The training with a diverse dataset significantly improves the model's generalizability. Through quantitative analysis, we have demonstrated the model's performance. Additionally, we developed a data transformation technique that can convert any velocity filtering problem into an upgoing and downgoing wave separation task. By combining the big data-trained model and data transformation technique, we successfully implemented deep learning velocity filtering. Through several practical application examples, we further validated the effectiveness and versatility of our proposed method.

Currently, data collection is limited. However, the method we proposed provides a new solution for velocity filtering. The data preparation process is both simple and effective, and with sufficient data sources, there is great potential to further improve the model's generalizability and accuracy.

DATA AND MATERIALS AVAILABILITY

The dataset and code associated with this study have been made publicly available at https://geolxb.github.io/openWFS/.

# REFERENCES

Abbad, B., B. Ursin, and M. J. Porsani, 2011, A fast, modified parabolic Radon transform: GEOPHYSICS, **76**, V11–V24.

Breuer, A., N. Ettrich, and P. Habelitz, 2020, Deep learning in seismic processing: Trim statics and demultiple, *in* SEG Technical Program Expanded Abstracts 2020, , . SEG Technical Program Expanded AbstractsSociety of Exploration Geophysicists, 3199–3203.

Cheng, S., Z. Cheng, C. Jiang, W. Mao, and Q. Zhang, 2024, An effective self-supervised learning method for attenuating various types of seismic noise: GEOPHYSICS, **89**, V589–V604.

Christie, P. a. F., V. J. Hughes, and B. L. N. Kennett, 1983, Velocity Filtering of Seismic Reflection Data: First Break, **1**.

Duncan, G., and G. Beresford, 1994, Slowness adaptive f-k filtering of prestack seismic data: GEOPHYSICS, **59**, 140–147.

Duncan, G., and G. Beresford, 1995a, Some analyses of 2-D median f-k filters: GEOPHYSICS, **60**, 1157–1168.

Duncan, G., and G. Beresford, 1995b, Median filter behaviour with seismic data: Geophysical Prospecting, **43**, 329–345.

Fernandez, M., N. Ettrich, M. Delescluse, A. Rabaute, and J. Keuper, 2024, Towards flexible demultiple with deep learning, *in* Fourth International Meeting for Applied Geoscience & Energy, , . SEG Technical Program Expanded AbstractsSociety of Exploration Geophysicists and American Association of Petroleum Geologists, 1958–1962.

Franco, de R., and G. Musacchio, 2001, Polarization filter with singular value decomposition: GEOPHYSICS, **66**, 932–938.

Freire, S. L. M., and T. J. Ulrych, 1988, Application of singular value decomposition to vertical seismic profiling: GEOPHYSICS, **53**, 778–785.

Guo, Y., S. Peng, W. Du, and D. Li, 2023, Denoising and wavefield separation method for DAS VSP via deep learning: Journal of Applied Geophysics, **210**, 104946.

Huo, J., B. Zhou, Q. Zhao, I. M. Mason, and Y. Rao, 2019, Migration-based filtering: Applications to geophysical imaging data: GEOPHYSICS, **84**, S219–S228.

Kaur, H., S. Fomel, and N. Pham, 2020, Seismic ground-roll noise attenuation using deep learning: Geophysical Prospecting, **68**, 2064–2077.

Kazemnia Kakhki, M., A. Mokhtari, and W. J. Mansur, 2024, Seismic data filtering using deconvolutive short-time Fourier transform: GEOPHYSICS, **89**, V243–V252.

Kommedal, J. H., and B. A. Tjøstheim, 1989, A Study of Different Methods of Wavefield Separation for Application to Vsp Data1: Geophysical Prospecting, **37**, 117–142.

Li, C., and J. Zhang, 2022, Wavefield separation using irreversible-migration filtering: GEOPHYSICS, **87**, A43–A48.

Li, X., Q. Qi, H. Huang, Y. Yang, P. Duan, and Z. Cao, 2023, An efficient deep learning method for VSP wavefield separation: A DAS-VSP case: GEOPHYSICS, **88**, WC91–WC105.

Li, X., Q. Qi, L. Li, P. Duan, Lin. L, J. Meng, and M. Zheng, 2025, A big data-driven deep learning method for seismic wavefield separation of VSP data: GEOPHYSICS (Accept Pending)

Lu, C., Z. Mu, J. Zong, and T. Wang, 2024, Unsupervised VSP Up- and Downgoing Wavefield Separation via Dual Convolutional Autoencoders: IEEE Transactions on Geoscience and Remote Sensing, **62**, 1–15.

Luo, Y., G. Zhang, J. Duan, C. Liang, Q. Wu, F. Yang, and X. Li, 2023, Two-Stage Multitask U-Network VSP Wavefield Separation: IEEE Geoscience and Remote Sensing Letters, **20**, 1–5.

March, D. W., and A. D. Bailey, 1983, A Review of the Two-dimensional Transform and its use in seismic processing: First Break, **1**.

Margrave, G. F., and P. F. Daley, 2014, VSP modelling in 1D with Q and buried source: , **26**.

Meng, T., T. Wang, J. Cheng, P. Duan, and Z. Cao, 2025, Up/down and P/S decomposition of DAS-VSP Data using multi-task deep learning method: GEOPHYSICS, 1–55.

Mitchell, A. R., and P. G. Kelamis, 1990, Efficient tau-p hyperbolic velocity filtering: GEOPHYSICS, **55**, 619–625.

Moon, W., A. Carswell, R. Tang, and C. Dilliston, 1986, Radon transform wave field separation for vertical seismic profiling data: GEOPHYSICS, **51**, 940–947.

Oliveira, D. A. B., D. G. Semin, and S. Zaytsev, 2021, Self-Supervised Ground-Roll Noise Attenuation Using Self-Labeling and Paired Data Synthesis: IEEE Transactions on Geoscience and Remote Sensing, **59**, 7147–7159.

Schonewille, M., and P. Zwartjes, 2002, High?resolution transforms and amplitude preservation, *in* SEG Technical Program Expanded Abstracts 2002, , . SEG Technical Program Expanded AbstractsSociety of Exploration Geophysicists, 2066–2069.

Stewart, R., 1985, Median filtering: Review and a new F/K analogue design: Journal of the Canadian Society of Exploration Geophysicists, **21**.

Tao, B., Y. Yang, H. Zhou, Y. Wang, F. Lyu, and W. Li, 2023, Deep Learning-based Upgoing and Downgoing Wavefield Separation for Vertical Seismic Profile Data: GEOPHYSICS, 1–80.

Trad, D., T. Ulrych, and M. Sacchi, 2003, Latest views of the sparse Radon transform: GEOPHYSICS, **68**, 386–399.

Trad, D. O., T. J. Ulrych, and M. D. Sacchi, 2002, Accurate interpolation with high-resolution time-variant Radon transforms: GEOPHYSICS, **67**, 644–656.

Wang, C., X. Huang, Y. Li, and K. Jensen, 2023, Removing multiple types of noise of distributed acoustic sensing seismic data using attention-guided denoising convolutional neural network: Frontiers in Earth Science, **10**.

Wen, Y., F. Qian, W. Guo, J. Zong, D. Peng, K. Chen, and G. Hu, 2025, VSP Upgoing and Downgoing Wavefield Separation: A Hybrid Model-Data-Driven Approach: IEEE Transactions on Geoscience and Remote Sensing, **63**, 1–14.

Willis, M. E., 2022, Distributed Acoustic Sensing for Seismic Measurements – What Geophysicists and Engineers Need to Know: Society of Exploration Geophysicists.

Yang, L., S. Wang, X. Chen, O. M. Saad, W. Cheng, and Y. Chen, 2023, Deep Learning with Fully Convolutional and Dense Connection Framework for Ground Roll Attenuation: Surveys in Geophysics, **44**, 1919–1952.

Yi, B., 2018, Weighted least squares Radon transform and its application in suppressing deep-formation multiple, *in* International Geophysical Conference, Beijing,

China, 24-27 April 2018, , . SEG Global Meeting AbstractsSociety of Exploration Geophysicists and Chinese Petroleum Society, 270–273.

Yu, S., J. Ma, and W. Wang, 2019, Deep learning for denoising: GEOPHYSICS, **84**, V333–V350.

Yuan, Y., X. Si, and Y. Zheng, 2020, Ground-roll attenuation using generative adversarial networks: GEOPHYSICS, **85**, WA255–WA267.